\documentstyle[12pt,epsfig]{article}
\def\Journal#1#2#3#4{{#1} {\bf #2}, #3 (#4)}

\def\NPA{{\em Nucl. Phys.} A}
\def\NPB{{\em Nucl. Phys.} B}
\def\PLB{{\em Phys. Lett.}  B}
\def\PRL{\em Phys. Rev. Lett.}
\def\PRD{{\em Phys. Rev.} D}
\def\ZPC{{\em Z. Phys.} C}

\def\beq{\begin{equation}}
\def\eeq{\end{equation}}
\def\lsim{\ ^<\llap{$_\sim$}\ }
\def\gsim{\ ^>\llap{$_\sim$}\ }
\def\r2{\sqrt 2}
\def\beq{\begin{equation}}
\def\eeq{\end{equation}}
\def\beqn{\begin{eqnarray}}
\def\eeqn{\end{eqnarray}}

\def\sinW2{\sin^2\theta_W}

\def\mz2{M_{z}^2}
\def\c2b{\cos 2\beta}

\def\mz{M_z}

\def\Fq2{F_{2}(q^2)}

\def\sec2w{sec^2\theta_W}
\def\amu{a_\mu}

\def\gmin2{(g-2)_\mu}
\def\ecoup{{e \over {{\sqrt2}\sin{\theta_W}}}}
\def\winomass{m_{\tilde W_a}}

\catcode`\@=11 

\def\lsim{\mathrel{\mathpalette\@versim<}}
\def\gsim{\mathrel{\mathpalette\@versim>}}
\def\@versim#1#2{\vcenter{\offinterlineskip
    \ialign{$\m@th#1\hfil##\hfil$\crcr#2\crcr\sim\crcr } }}
\def\etal{{\em et. al.}}

\def\PRL{Phys. Rev. Lett.}

\begin{document}
\begin{flushright}
{CERN-TH/2001-232}\\
{TIFR/TH/01-33}\\ 
\end{flushright}
\begin{center}
{\Large\bf Muon g-2 and Implications for Supersymmetry\\}
\vglue 0.5cm
{Utpal Chattopadhyay$^{(a)}$ and 
Pran Nath$^{(b,c)}$
\vglue 0.2cm
{\em 
$^a$Department of Theoretical Physics,\\ Tata Institute
of Fundamental Research,Homi Bhabha Road\\
Mumbai 400005, India}\\
{\em $^{(b)}$ Theoretical Physics Division, CERN CH 1211, Geneva, 
Switzerland\\}
{\em $^{(c)}$Department of Physics, Northeastern University, Boston,
MA 02115, USA\footnote{Permanent address}\\} }
\end{center}
\begin{abstract}
A brief review is given of the implications of the recent Brookhaven result
on the muon anomaly ( $a_{\mu}$) for supersymmetry. We focus mainly on the 
implications of the recent results for the minimal supergravity 
unified model. We show that the observed difference implies the 
existence of sparticles  most of which should become observable
at the Large Hadron Collider. Further, as foreseen in works prior to
the Brookhaven experiment the sign of the  difference between
 experimental prediction of $a_{\mu}$ and its Standard Model value
 determines the sign of the Higgs mixing parameter $\mu$. 
The $\mu$ sign has important implications for the direct detection
of dark matter. Implications of the Brookhaven result for other 
low energy phenomena are also discussed. 
\end{abstract}

\section{Introduction}
In this talk we give a brief discussion of the recent 
developments in the analyses of the muon anomaly.
First, we will discuss
the  recent Brookhaven National Laboratory (BNL) 
result on $a_{\mu}$\cite{brown} ($a=(g-2)/2$ where g is the gyromagnetic
ratio) and its Standard Model prediction.
We then discuss the supersymmetric electro-weak effects on
  $a_{\mu}$.  We will also discuss briefly the effects of extra 
  dimensions on $a_{\mu}$.  Finally we will discuss 
  the implications of the BNL 
 result for the direct detection of 
 supersymmetry. The anomalous moment is a sensitive probe of new 
physics since 

\begin{equation}
a^{new -physics}_l \sim \frac{m_l^2}{\Lambda^2}
\end{equation}
Thus $a_{\mu}$ is more sensitive to new physics relative to $a_e$ 
even though $a_e$ is more accurately determined\cite{dehmelt}
   since $\frac{a_{\mu}}{a_e}\sim 4\times 10^4$.
 Regarding the experimental determination of $a_{\mu}$ one has first 
 the classic CERN experiment of 1977\cite{bailey} which gave 
$a_{\mu}^{exp}=11659230(84)\times 10^{-10}$. The error in this measurement 
was reduced by a factor of 2 in 1998 by the BNL 
experiment\cite{bnl98} which gave 
$a_{\mu}^{exp}=11659205(46)\times 10^{-10}$, and the same error
was further reduced by a factor of 3 by the most recent BNL 
result\cite{brown}, i.e.,

\begin{equation}
a_{\mu}^{exp}=11659203(15)\times 10^{-10}
\end{equation}
The  Standard Model contribution consists of several parts\cite{czar1}

\begin{equation}
a_{\mu}^{SM}=a_{\mu}^{qed}+a_{\mu}^{EW}+a_{\mu}^{hadronic}
\end{equation}
where the qed correction is computed to order $\alpha^5$\cite{czar1}
\begin{equation}
a_{\mu}^{qed}=11658470.57(.29)\times 10^{-10}
\end{equation}
and $a_{\mu}^{EW}$ including the one loop\cite{fuji} and the 
two loop\cite{czar1} Standard Model electro-weak correction is 
\begin{equation}
a_{\mu}^{EW}=15.2(0.4)\times 10^{-10}
\end{equation}
The most difficult
part of the analysis relates to the hadronic contribution. 
It consists of several
parts: the $\alpha^2$ hadronic vacuum polarization contribution,
the $\alpha^3$ hadronic correction, and the light-by-light 
contribution.
The $\alpha^2$ hadronic vacuum polarization contribution can be
related to observables. Specifically one can write

\begin{equation}
a^{had}_{\mu}(vac.pol.)=(\frac{1}{4\pi^3})
\int_{4m_{\pi}^2}^{\infty}{ds} K(s) \sigma_h(s)
\end{equation}
where $\sigma_h(s)=\sigma (e^+e^-\rightarrow hadrons)$
and $K(s)$ is a kinematical factor.
The integral in Eq.(6) is dominated by the low energy part,
i.e., the part up to 2 GeV,
which correspondingly is also very sensitive to errors 
in the input data. In the evaluations of Eq.(6) one uses a combination
 of experimental 
data at low energy and a theoretical (QCD) extrapolation in the
high energy tail. The analysis of 
$a^{had}_{\mu}(vac.pol.)$ is the most contentious part of the 
analysis.
 In computing the difference $a_{\mu}^{exp}-a_{\mu}^{SM}$, 
BNL used the result of Davier and Hoker\cite{davier}, i.e., 
 $a_{\mu}^{had}(vac.pol.)=692.4 (6.2)\times 10^{-10}$.  
 However, other estimates have appeared more recently and we will
 mention these later. The $\alpha^3$ hadronic correction 
 can also be related to  observables but is generally small 
 with a correspondingly small error\cite{krause}, i.e., 
$\Delta a_{\mu}^{had}(vac.pol.)=-10.1(.6)\times 10^{-10}$.
The light-by-light hadronic correction is the second most 
contentious part of $a_{\mu}^{SM}$. This part cannot
be related to any observables and is thus a purely theoretical
construct. In the free quark model it evaluates to a positive
contribution. However, more realistic analyses give a negative
contribution\cite{bijnens}, i.e., 
$\Delta a_{\mu}^{had}(light-by-light)=-8.5(2.5)\times 10^{-10}$.
This result which is though more reliable than the result from the free 
quark model, still has a degree of model dependence.   
Overall, however,  $\Delta a_{\mu}^{had}(light-by-light)$ 
is not the controlling factor in
interpreting the BNL result unless, of course, its sign
is reversed.  The total result then is
$ a_{\mu}^{had}(total)=a_{\mu}^{had}(vac.pol.)
+ \Delta a_{\mu}^{had}(vac.pol.) +\Delta a_{\mu}^{had}(light-by-light)$
which gives

\begin{equation}
a_{\mu}^{hadronic}=673.9(6.7)\times 10^{-10}
\end{equation}
Together one finds,    
\begin{equation}
a_{\mu}^{SM}= 11659159.7(6.7)\times 10^{-10}
\end{equation}
and a $2.6$ sigma deviation of experiment from theory,

\begin{equation}
a_{\mu}^{exp}- a_{\mu}^{SM}=43(16)\times 10^{-10} . 
\end{equation}
After the new $g-2$ result from Brookhaven became available,
there have been  several reanalyses of the hadronic
uncertainty\cite{yndurain,narison,melnikov,cvetic,mr}.  
Thus, e.g., the analysis of Ref.\cite{narison} gives
$\Delta= 33.3(17.1)$   and  of Ref.\cite{melnikov}
 gives $\Delta= (37.7\pm (15.0)_{exp}\pm (15.6)_{th})$
 where $\Delta=(a_{\mu}^{exp}- a_{\mu}^{SM})\times 10^{10}$.
 One finds that the difference  $(a_{\mu}^{exp}- a_{\mu}^{SM})$
 in these analyses is somewhat smaller and the error 
 somewhat larger compared to the result of Eq.(9). Similar trends
 are reported in the analyses of Refs.\cite{yndurain,cvetic}.
 An interesting assessment of the hadronic contribution and 
 the possibilities for improvement in the future is given in
 Ref.\cite{mr}. For the discussion of the rest of this paper 
 we will assume the validity of Eq.(9).
 
One may ask what is the nature of new physics in view of Eq.(9).
 Some possibilities 
that present themselves are supersymmetry, compact extra dimensions,
muon compositeness, technicolor, anomalous W couplings, 
new gauge bosons, lepto-quarks and radiative muon masses.
We shall focus here mostly on supersymmetry as the possible 
origin of the difference observed by the BNL experiment.
Supersymmetry has many attractive features. It helps to stabilize
the hierarchy problem with fundamental Higgs, and it leads to the 
unification of the gauge coupling constants consistent with the LEP data. 
To extract meaningful results from SUSY models, however, one needs a  
mechanism of supersymmetry breaking. There are several mechanisms
proposed for the breaking of supersymmetry such as gravity mediated, gauge
mediated, anomaly mediated etc. We focus in this paper mainly
on the gravity mediated models, i.e., the supergravity (SUGRA) unified 
models\cite{chams}. In the
minimal version of this model based on a flat K\"ahler 
potential, ie.,  mSUGRA, 
 the SUSY breaking sector is described by the parameters $m_0$, 
 $m_{\frac{1}{2}}$, $A_0$, $\tan\beta$ and $sign(\mu)$  
 where $m_0$ is the universal scalar mass,  $m_{\frac{1}{2}}$
 is the universal gaugino mass, $A_0$ is the universal trilinear
 coupling,  $\tan\beta =<H_2>/<H_1>$ where $H_2$ gives mass to
 the up quark and $H_1$ gives mass to the down quark and the lepton,
  and $\mu$ is the Higgs mixing parameter. The use of the 
   curved K\"ahler potential results in  a SUGRA model with non-universalities
    consisting of the  minimal set of soft SUSY parameters and additional
    parameters which, for example, describe deviations from universality
    in the Higgs sector and in the  third generation sector.
    
    Some of the interesting features of SUGRA models include the fact that
    the radiative breaking  constraints of the electro-weak symmetry
    leads to the lightest neutralino being the lightest 
    supersymmetric particle (LSP) and thus under the constraint of R parity
    the lightest neutralino is a possible candidate for dark matter
    over most of the parameter space of the model. 
    Also, analyses in SUGRA
    models show that the lightest Higgs must have a mass
    $m_h\leq 130$ GeV under the usual assumptions of naturalness, i.e.,
    $m_0, m_{\tilde g}<1$~TeV. Finally, SUGRA models bring in new
    sources of CP violation which in any case are needed for 
    baryogenesis. Thus, mSUGRA has two soft CP violating phases
    while, many more soft CP violating phases arise in non-universal 
    SUGRA models and in the minimal supersymmetric standard model 
    (MSSM). Regarding some of the other alternatives 
    of SUSY breaking, one  finds  that the gauge mediated breaking
    (GMSB) does not produce a candidate for cold dark matter,  
    while the anomaly mediated supersymmetry breaking (AMSB) 
    scenario now appears very stringently constrained
    by the BNL data when combined with its specially characteristic
    $b \rightarrow  s + \gamma$ constraint.    

Our analysis in mSUGRA includes two loop renormalization group 
evolutions (RGE) for the couplings as well as soft parameters 
with the Higgs potential at the complete one-loop level~\cite{oneloopeff} 
minimized at the scale $Q \sim \sqrt{m_{{\tilde t}_1}m_{{\tilde t}_2}}$
for radiative electroweak symmetry breaking. We have also included the SUSY 
QCD corrections~\cite{susybtmass} 
to the top quark (with $M_t=175$ GeV) and the bottom quark 
masses and we have used the code {\it FeynHiggsFast}~\cite{feynhiggs}
for the mass of the light Higgs boson.  


\section{ SUSY contribution to $a_\mu$ at one loop}
 It is well known that $a_{\mu}$ vanishes in the exact supersymmetric
 limit\cite{ferrara} and is non-vanishing only in the presence of 
 supersymmetry breaking. Not surprisingly then $a_{\mu}^{SUSY}$ (where 
$a_{\mu}=a_{\mu}^{SM}+a_{\mu}^{SUSY}$) is sensitive to the nature
 of new physics\cite{fayet}. Thus the analysis of $a_{\mu}^{SUSY}$
 requires a realistic model of supersymmetry breaking.
  The first such analysis within 
 the well motivated SUGRA  model was given in 
 Refs.\cite{kosower,yuan}. We reproduce here
 partially the result of Ref.\cite{yuan}

\begin{equation}
a_{\mu}^{SUSY}={a_{\mu}^{\tilde W}+a_{\mu}^{\tilde Z}}
\end{equation}
where $a_{\mu}^{\tilde W}$ is the chargino contribution
and $a_{\mu}^{\tilde Z}$ is the neutralino contribution.
The chargino contribution is typically the larger contribution 
over most of the parameter space  and is 

\begin{equation}
 a_{\mu}^{\tilde W}={{m^2_\mu} \over {48{\pi}^2}} {{ {A_R^{(a)}}^2} \over
{\winomass^2}}F_1\left(({{m_{{\tilde \nu}_\mu}} \over {m_{\tilde W_a}}})^2
\right)+
{{m_\mu} \over{8{\pi}^2}} {{A_R^{(a)}
A_L^{(a)}} \over {\winomass}} 
F_2\left(({{m_{{\tilde \nu}_\mu}} \over {m_{\tilde W_a}}})^2\right)
\end{equation}
Here $A_L(A_R)$ are the left(right) chiral amplitudes 

\begin{equation}
A_R^{(1)}=-{\ecoup}\cos {\gamma_1}; \quad A_L^{(1)}={(-1)^\theta}
{{{em_\mu}\cos{\gamma_2}} \over {{2}M_W\sin\theta_W \cos {\beta}}}
\end{equation}

\begin{equation}
A_R^{(2)}=-{\ecoup}\sin {\gamma_1}; \quad A_L^{(2)}=-
{{{e m_\mu}\sin{\gamma_2}} \over
{{2}M_W \sin\theta_W\cos {\beta}}}
\end{equation}
where $\theta =0(1)$ if the light chargino eigenvalue 
$\lambda_1$ is positive (negative), and $\gamma_{1,2}$ are 
mixing angles.
We wish to point out that the most dominant contribution to 
$a_\mu^{SUSY}$ comes from the chirality non-diagonal lighter chargino 
part of $a_{\mu}^{\tilde W}$.  
First we note that for the most contributing term in the chargino part 
the coupling is proportional to $1/\cos\beta (\sim \tan \beta)$ and thus 
 $a_{\mu}$ increases almost linearly with $\tan\beta$ 
\cite{lopez,chatto};
second due to the same dominant term the sign of 
$a_{\mu}^{SUSY}$ is correlated strongly with the 
sign of $\mu$ (we use here the $\mu$ sign convention of Ref.\cite{sugra}).
It is easy to exhibit this by considering the eigenvalues $\lambda_i$
(i=1,2) of the chargino mass matrix (where we define $\lambda_1$
as the eigenvalue corresponding to the lighter chargino) that
 $\lambda_1 <0$ for $\mu>0$ and  $\lambda_1 >0$ for $\mu<0$
except  for $\tan\beta \sim 1$, which leads to\cite{lopez,chatto}
 $\amu^{SUSY}>0, \mu>0$ and $\amu^{SUSY}<0, \mu<0$.

\section{Implications of Precise BNL Data}
In the following analysis we assume CP conservation.
Under this constraint and setting
$a_{\mu}^{SUSY}=a_{\mu}^{exp}-a_{\mu}^{SM}$, we immediately
find that the BNL data determines\cite{chatto2,kane,feng,gondolo}
 $sign(\mu)=+1$.  In imposing the BNL constraint we use  a 
 $2\sigma$ corridor
\begin{equation}
 10.6\times 10^{-10} <a_{\mu}^{SUSY}<76.2\times 10^{-10}
\end{equation}
\begin{figure}[hbt]
\centerline{\epsfig{file=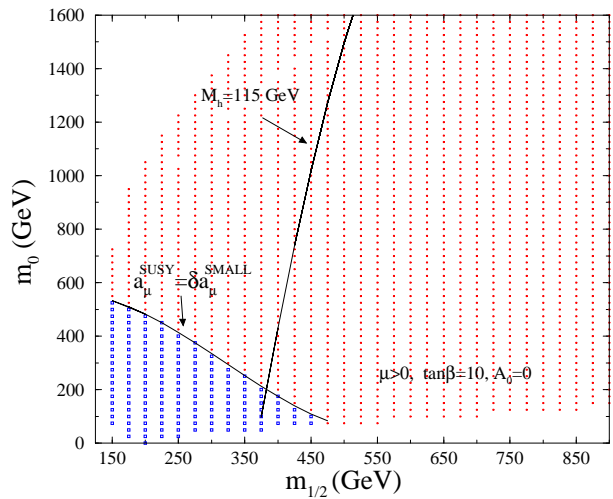,width=0.50\textwidth}}
\caption{Upper limit in the $m_0-m_{\frac{1}{2}}$ plane implied by
the BNL $g-2$ constraint for $\tan\beta =10$ indicated by the line 
$a_{\mu}^{SUSY}=\delta a_{\mu}^{SMALL}=10.6\times 10^{-10}$. 
The allowed region in the
parameter consistent with constraint of Eq.(14) lies below this line.
 115 GeV Higgs signal\cite{lep} is also indicated (from Ref.\cite{chatto2}).
}
\label{tan10}
\end{figure}

We utilize Eq.(14) in determining the allowed parameter space of
mSUGRA using the one loop formula for which the chargino part is 
given by Eq.(11)\cite{yuan}.
[The leading order 
correction to one loop as computed in Ref.\cite{giudice} gives a 
fractional contribution of $-(4\alpha/\pi)ln(M_S/m_{\mu})$ where
$M_S$ is an average sparticle mass. This is typically less than $10\%$
and is ignored in the analysis here.]  In Fig.\ref{tan10} we give 
an analysis of this constraint in the $m_0-m_{\frac{1}{2}}$
plane for the case of $\tan\beta =10$. One finds that there is
now an upper limit on $m_0$ and  $m_{\frac{1}{2}}$.
Interestingly, we find that the allowed region of the parameter 
space which is below the $a_{\mu}^{SUSY}=\delta a_{\mu}^{SMALL}= 
10.6\times 10^{-10}$
line allows for a light Higgs consistent with the lower limit
of about 115 GeV as given by the possible signal at LEP\cite{lep}.
The white region close to $m_{\frac{1}{2}}$ axis in 
Fig.\ref{tan10} is excluded 
for stau turning to be the LSP. 
The left side white region near the $m_0$ axis is excluded by 
the constraints from chargino mass lower limit or radiative 
electro-weak symmetry breaking.
\begin{figure}[hbt]
\centerline{\epsfig{file=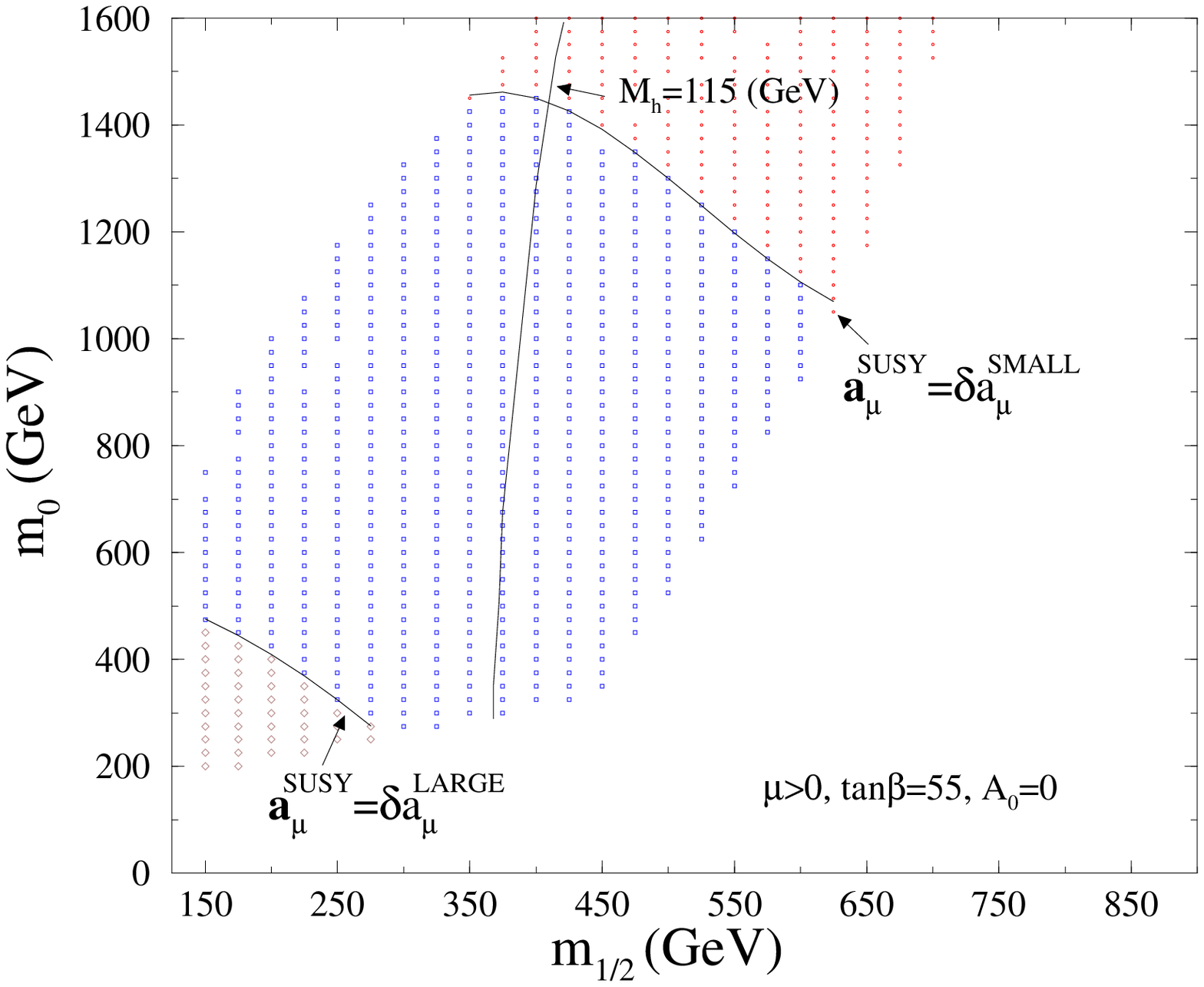,width=0.50\textwidth}}
\caption{Upper and lower limits in the $m_0-m_{\frac{1}{2}}$ plane 
implied by the BNL $g-2$ constraint for $\tan\beta$ indicated by lines 
$a_{\mu}^{SUSY}=\delta a_{\mu}^{SMALL}=10.6\times 10^{-10}$ and 
$a_{\mu}^{SUSY}=\delta a_{\mu}^{LARGE}=76.2\times 10^{-10}$. 
 The allowed region in the
parameter consistent with constraint of Eq.(14) lies between the lines.
 The 115 GeV Higgs signal is also indicated (from Ref.\cite{chatto2}). 
}
\label{tan55}
\end{figure}

Next we discuss the case of a large $\tan\beta$, i.e., 
$\tan\beta =55$. This is the largest $\tan\beta$ before one
gets into a non-perturbative domain for most of the parameter space. 
The results of the analysis on the allowed parameter 
space in the $m_0-m_{\frac{1}{2}}$ plane
are given in Fig.\ref{tan55} consistent with the constraints of Eq.(14). 
One finds that  in this case there is both a lower limit 
and an upper limit and the allowed parameter space is the 
shaded area contained between the lines. The white region near 
$m_0$ axis for larger $m_0$ and smaller $m_{\frac{1}{2}}$ values 
is excluded because of the chargino mass lower limit or 
the radiative electro-weak symmetry breaking constraints. The white 
region near $m_{\frac{1}{2}}$ axis having smaller $m_0$ values 
is excluded via the tachyonic stau constraint and the region just above this, 
corresponding to moderately large $m_0$ values is excluded because of the 
CP-odd Higgs boson turning tachyonic at the tree level which is a large 
$\tan\beta$ effect.   
Again the Higgs signal corresponding to the LEP lower limit is indicated 
by the solid near vertical line and is seen to lie in the 
allowed region of the parameter space.

\begin{figure}[hbt]
\centerline{\epsfig{file=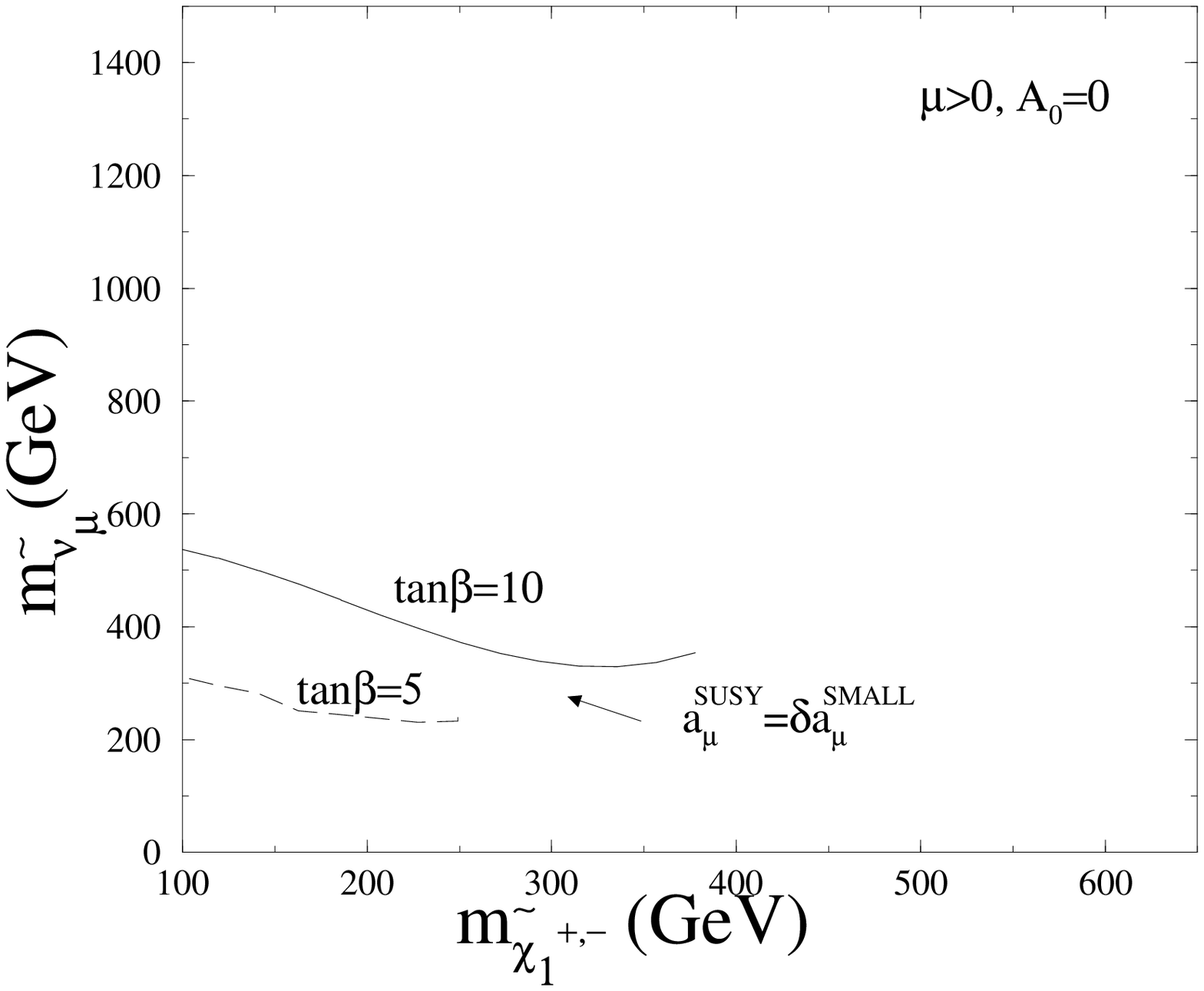, width=0.50\textwidth}}
\caption{Upper limits in the 
$m_{\tilde \nu_{\mu}}-m_{\tilde \chi_1^{+,-}}$  plane 
implied by the BNL $g-2$ constraint for $\tan\beta =5$ and 10,  
 are indicated by the lines $a_{\mu}^{SUSY}=\delta a_{\mu}^{SMALL}
=10.6\times 10^{-10}$.
 The allowed region in the
parameter consistent with constraint of Eq.(14) lies below the 
lines (from Ref.\cite{chatto2}).
}
\label{ch_5_10}
\end{figure}

\begin{figure}[hbt]
\centerline{\epsfig{file=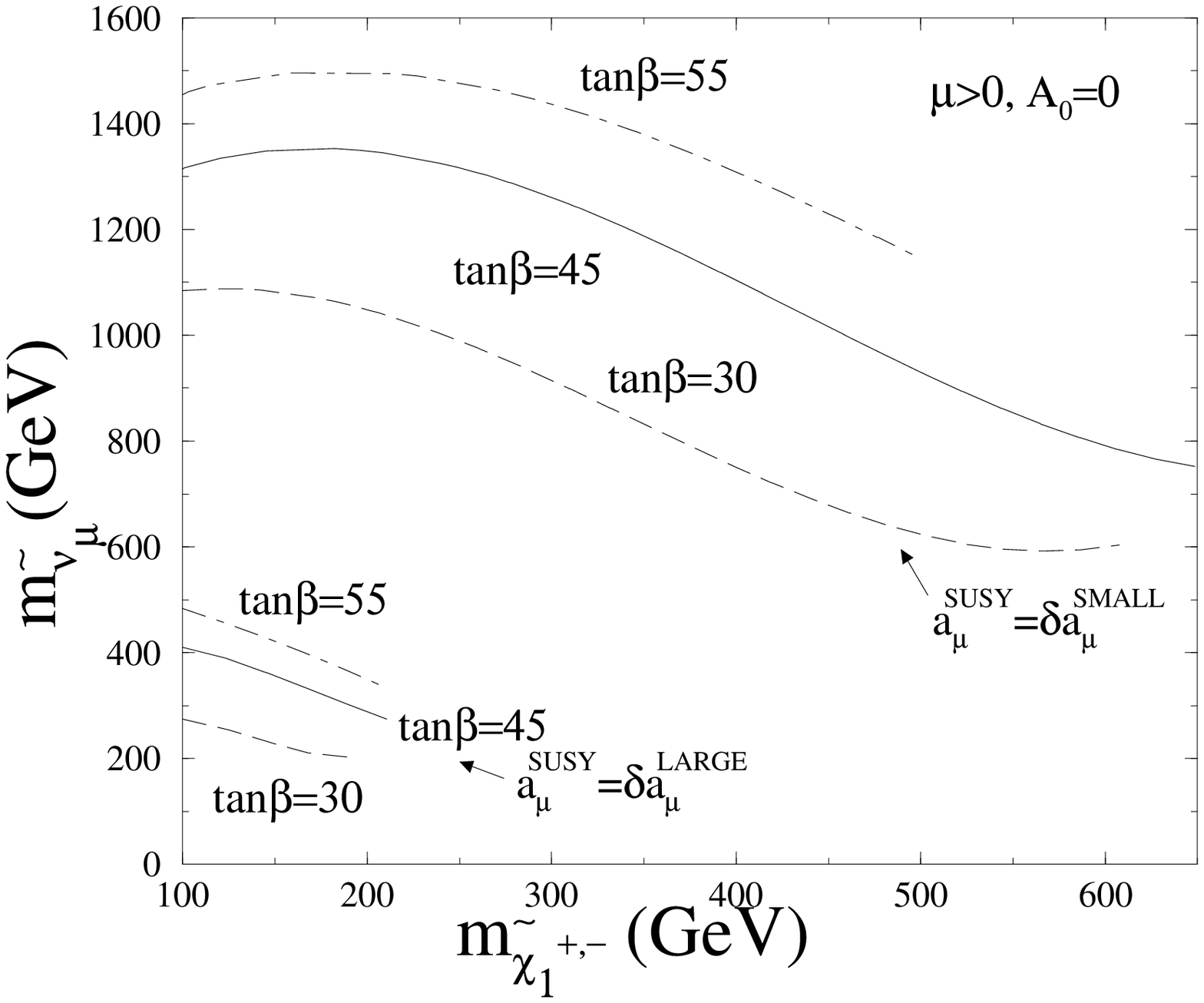, width=0.50\textwidth}}
\caption{Upper limits and lower limits in the 
$m_{\tilde \nu_{\mu}}-m_{\tilde \chi_1^{+,-}}$  plane 
implied by the BNL $g-2$ constraint for $\tan\beta =30,45$ and 55 
 are indicated by lines  $a_{\mu}^{SUSY}=\delta a_{\mu}^{SMALL}
=10.6\times 10^{-10}$
 and $a_{\mu}^{SUSY}=\delta a_{\mu}^{LARGE} =76.2\times 10^{-10}$.
 The allowed region in the
parameter consistent with constraint of Eq.(14) lies between the 
lines (from Ref.\cite{chatto2}).
}
\label{ch_30_45_55}
\end{figure}

A full analysis was carried out including also values of
 $\tan\beta =5,30$ and $45$ in Ref.\cite{chatto2}. 
 We discuss the results of the full analysis from the point of view 
 of sparticle spectra. In Fig.\ref{ch_5_10} the upper 
limits in sneutrino-light chargino
plane are given for $\tan\beta =5$ and $10$.  A similar analysis is given 
for $\tan\beta = 30, 45$, and $55$ in Fig.\ref{ch_30_45_55}.  
From Fig.\ref{ch_5_10} and Fig.\ref{ch_30_45_55} one finds,
as expected, that there are strong correlations between the upper  
limits and $\tan\beta$. Using the entire data set in Fig.\ref{ch_5_10} 
and Fig.\ref{ch_30_45_55} one finds,

 \begin{equation}
m_{{\tilde \chi}_1^\pm}\leq 650 {~\rm GeV}, m_{{\tilde \nu}_\mu}\leq 1.5 {~\rm TeV} 
~(\tan\beta\leq 55)
\end{equation}
The corresponding limits in the $m_0-m_{\frac{1}{2}}$ plane are   

\begin{equation}
m_{1/2}\leq 800 {~\rm GeV}, m_0\leq 1.5 {~\rm TeV} ~(\tan\beta\leq 55)
\end{equation}

\begin{figure}[hbt]
\centerline{\epsfig{file=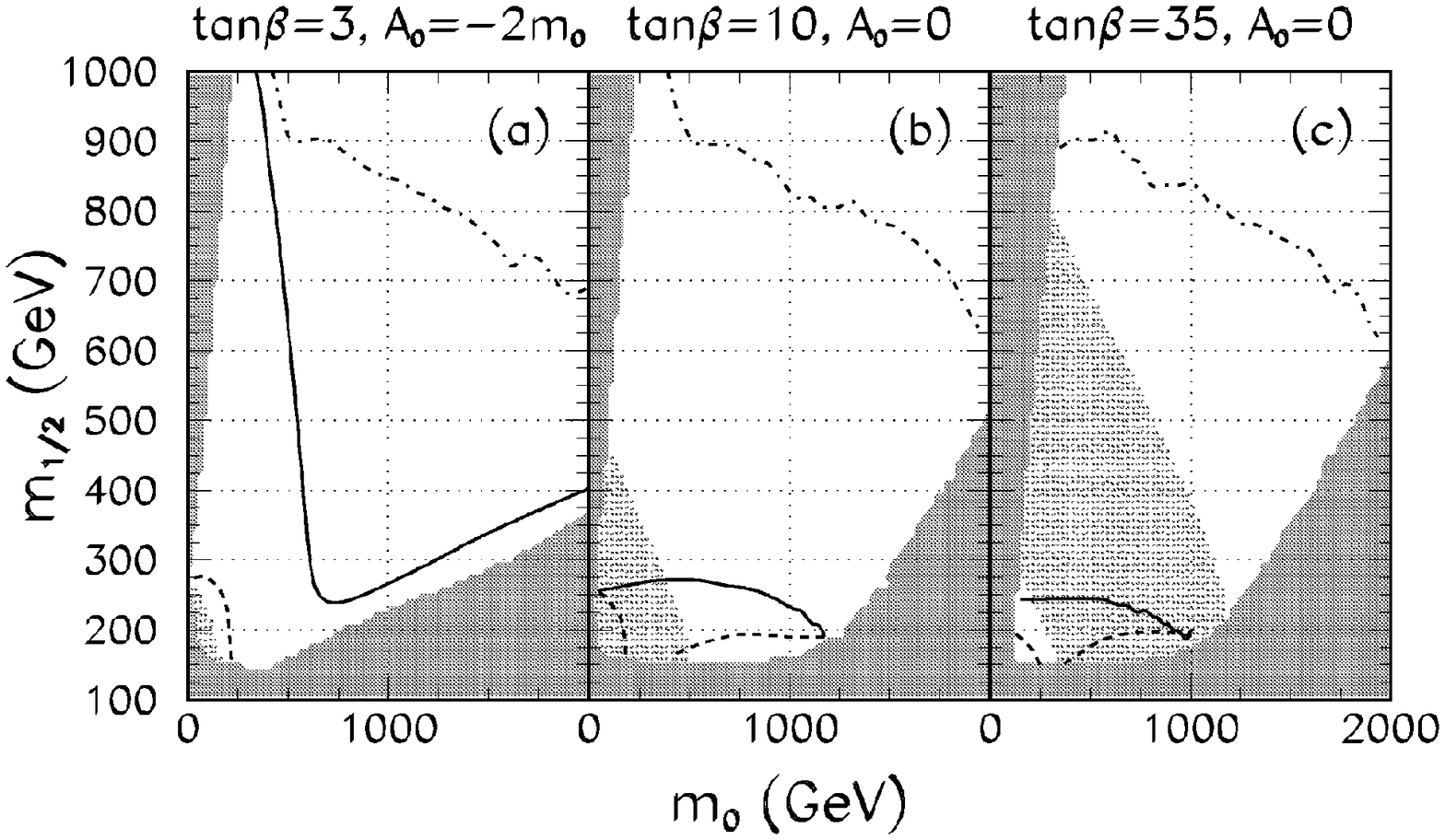,width=0.90\textwidth}}
\caption{A plot of $m_0\ vs.\ m_{1/2}$ parameter space 
in the mSUGRA model
for $\mu >0$ and {\it a}) $A_0=-2m_0$ and $\tan\beta =3$, {\it b}) $A_0=0$ and
$\tan\beta =10$ and {\it c}) $A_0=0$ and $\tan\beta =35$. The $2\sigma$
region favored by the E821 measurement is shaded with dots. The region
below the
solid contour has $m_h<113.5$ GeV. The region below the dashed
contour is accessible to Tevatron searches with 25 fb$^{-1}$ of integrated
luminosity, while the region below the dot-dashed contour is accessible via
LHC sparticle searches with 10 fb $^{-1}$ of integrated luminosity.
(Taken from Ref.\cite{baer}).
}
\label{mnewsugra}
\end{figure}

The upper limits that arise in mSUGRA from  the analysis of 
Ref.\cite{chatto2}  are consistent
with the fine tuning criteria (see, e.g., Ref.\cite{ccn}), and are very
encouraging from the point of view of discovery of superparticles
at colliders. Thus  LHC can discover squarks and gluinos up
to 2 TeV\cite{cms,baer}.  This means that essentially all of 
the squark and gluino mass spectrum  allowed within mSUGRA by the 
Brookhaven $g-2$ constraint will become 
visible at LHC\cite{chatto2,baer}. A comparison of the upper limits 
in the $m_0-m_{\frac[1]{2}}$ plane allowed by the g-2 constraint vs
the discovery potential of the LHC is given in the work of Baer et.al.
in Ref.\cite{baer} and we reproduce one
of the figures from that analysis here(see Fig.\ref{mnewsugra}).

Many further investigations of the implications of the BNL result
have been carried out over the recent 
months\cite{marciano,moroi,ellis2,adhs,wells}
exploring the effects of the $g-2$ constraint on a variety of low 
energy  phenomena such as on $b\rightarrow s+\gamma$,  dark matter,
 lepton flavor violation, trileptonic signal\cite{trilep}
 and on other low energy SUSY signals.  We briefly discuss two 
 of these: $b\rightarrow s+\gamma$ and dark matter.
Regarding $b\rightarrow s+\gamma$, the Standard Model branching ratio
for this process is estimated to be\cite{sm} $B(b\rightarrow s+\gamma)$$=$ 
$(3.29\pm  0.33)$$\times 10^{-4}$.
 Recent experiment gives\cite{cleo} 
$B(b\rightarrow s+\gamma)$$=$$ (3.15\pm 0.35\pm 0.32\pm 0.26)$
$\times 10^{-4}$ where the first error is statistical, 
and there are two types of systematic errors.
Now it is well known that the imposition
of the $b\rightarrow s+\gamma$ constraint puts severe limits on the
mSUGRA parameter space when $\mu<0$ eliminating most of the parameter
space in this case\cite{bsgamma,bsgammanew}. 
Thus had the sign of $\mu$ from the BNL experiment
turned out to be negative it would have eliminated most of the 
parameter space of the minimal model. On the other hand for the
 $\mu>0$ case one finds that  
 the constraint $b\rightarrow s+\gamma$ is much less
severe. Thus most of the parameter space of mSUGRA in this case 
at least for small and moderately large values of $\tan\beta$ is 
left  unconstrained. For large values of $\tan\beta$ nearing 50 the
$b\rightarrow s+\gamma$ constraint does become more stringent but
 a significant part of the parameter space is still allowed\cite{ellis2}.
However, it has been emphasized in Ref.\cite{wells} that 
$B(b\rightarrow s+\gamma)$  is not a pure observable and requires 
hard cuts for its extraction experimentally. This provides a note
of caution on imposing the $B(b\rightarrow s+\gamma)$ constraint
too stringently.

A closely related phenomenon that is sensitive to the sign of 
$\mu$ is the analysis of dark matter. It was shown in the
early days when the first measurement of $b\rightarrow s+\gamma$ was
made that the $b\rightarrow s+\gamma$ branching ratio has a strong
correlation with the neutralino-proton cross-sections in the direct detection 
of dark matter\cite{bsgamma} in regard to the sign of $\mu$. 
This happens due to the fact that the neutralino-proton 
cross-sections are smaller for the case of $\mu<0$ than for 
the case of $\mu>0$.
Additionally, with the $b\rightarrow s+\gamma$ constraint 
which eliminates most of the parameter space for 
$\mu<0$, one finds that the neutralino-proton cross 
sections to be very small for the available region of parameters 
for this sign of $\mu$.
Consequently, direct detection of neutralino dark matter is strongly 
disfavored for $\mu<0$ as opposed to what one finds for $\mu>0$. Thus,  
the fact that the BNL experiment determines 
the $\mu$ sign to be positive is indeed  good news for the direct 
detection of dark matter\cite{chatto2,gondolo,ellis2,adhs}.

We now turn to a brief discussion of  models other than mSUGRA.
One such model is AMSB.  The details of this model and procedure for its
implementation can be found in  Ref.\cite{chatto3}. The analysis 
for this case is given in Ref.\cite{chatto2} 
where the upper limits
in the  sneutrino-chargino plane corresponding 
to three values of $\tan\beta$, i.e., $\tan\beta =10, 30$, and
40 (the maximum allowed) were analyzed which produced upper limits of  
$m_{\tilde \nu_{\mu}}\leq 1.1 {~\rm TeV}$  
and $m_{\chi_1^+}\leq 300 {~\rm GeV}$. These limits are lower than those of
Eq.(15). 
Further, for $\mu>0$, 
one finds that the constraint from $b\rightarrow s+\gamma$ in this case
 excludes a significant amount of parameter space when the BNL 
$g-2$ constraint is imposed\cite{feng}. 
Further, analyses within the framework
of the unconstrained supersymmetric standard model, 
and analyses within more
general scenarios and their implications for colliders 
are given in works of Refs.\cite{moroi,kolda}.



One possibility which must be discussed along with supersymmetry is that
of  contributions from extra space time dimensions to $g-2$.
 In Ref.\cite{nyg2} a class of realistic 
models with extra spacetime dimensions were 
considered (For reviews see Refs.\cite{extrareview}). It was 
shown that for the case of one extra dimension compactified on
$S^1/Z_2$ with matter and Higgs fields residing on the orbifolds
and the gauge fields propagating in the bulk, the massless spectrum
of the model coincides with the massless spectrum of MSSM.
The Kaluza-Klein modes for W contribute to the Fermi constant
and the current good agreement between the Standard Model
determination of $G_F$ and its experimental value leaves only a small
error corridor in which the contributions from extra dimensions  
can reside. This constraint leads to a lower limit of about 3 TeV 
on the inverse compactified dimension and severely 
constrains the contribution of extra dimensions to the muon
anomalous magnetic moment. One finds that for the case of one extra dimension, 
the contribution of Kaluza-Klein states is smaller than the
supersymmetric contribution by more than two orders of magnitude.
For the case of more than one extra dimension, the contribution
to $a_{\mu}$ is larger than for the case of one extra dimension
but still significantly smaller than the one arising from 
supersymmetry. Thus we conclude that  models with
 extra dimensions of the type considered in Ref.\cite{nyg2}
 do not create a strong background  relative to the   
 supersymmetric effects (see, however, the analysis of Ref.\cite{desh}).
 
\section{Conclusions}
In this review we have given a brief summary of the developments
on the analyses of the muon anomaly. Implications of the difference
$a_{\mu}^{exp}-a_{\mu}^{SM}$ seen at BNL for supersymmetric
models and specifically for mSUGRA were explored. An effect of the
size seen at brookhaven for $a_{\mu}^{exp}-a_{\mu}^{SM}$ 
was already predicted within the SUGRA model in 1984
where it was found that the supersymmetric correction could be as large 
or larger than the Standard Model electro-weak correction\cite{yuan}.
Furthermore, we have also explored the implications of the
BNL result for the direct detection of supersymmetry at accelerators
and in dark matter searches. 
Thus a detailed analysis  within  mSUGRA of the BNL result using a $2\sigma$
error corridor on the difference $a_{\mu}^{exp}-a_{\mu}^{SM}$ 
leads to upper limits on sparticle masses which all lie below 2 TeV.
 Since the LHC can discover squarks and the gluino up to 2 TeV,  
 most if not all of the sparticles should become visible at LHC.
 Further, it was pointed out that the BNL data determines the 
 sign of $\mu$ to be positive within the minimal model which is very 
 encouraging for direct dark matter searches. 
It was also pointed out that there is little chance of confusing
the supersymmetric contribution to $a_{\mu}$ with effects from 
extra dimensions. This is so at least in models where the Standard
Model is obtained by a direct compactification of a five dimensional
model on $S^1/Z_2$  which gives a contribution to $a_{\mu}$ from 
Kaluza-Klein excitations, significantly smaller than 
a typical supersymmetric contribution. 
The BNL data also imposes impressive constraints on 
CP phases. It was shown in Ref.\cite{icn} that the BNL constraint eliminates
up to 60-90\% of the parameter space in the $\theta_{\mu}$ 
and $\xi_2$ (phase of ${\tilde m}_2 $) plane. In the presence of phases 
the relationship between the sign of $a_{\mu}^{SUSY}$ and 
the phase of $\mu$ may also be modified.

 There is a significant amount of data from the run of 2000 which 
would be analyzed
in the near future and BNL eventually hopes to measure 
$a_{\mu}$ to an accuracy of $4\times 10^{-10}$.
 Analyses including data from Beijing\cite{beijingref}, 
from Novosibirsk\cite{novoref} and additional
 $\tau$ data from CLEO\cite{cleoref} should 
delineate the hadronic error more reliably.
Further if deviation between theory and 
experiment persists at the current level after the analysis of the
 new data currently underway is carried out, 
and if also the error corridor 
shrinks then a signal for new physics will be undeniable. 
Such a signal interpreted as arising
from supersymmetry then has dramatic new predictions for the direct
observation of sparticles at accelerators.
Further, if supersymmetry is the right explanation for such
an effect, and there is a great bulk of theoretical reasoning 
in justification of this expectation, then  
 the search for a fundamental Higgs boson 
 becomes all the more urgent. Thus, the Brookhaven $g-2$ result
  further heightens the 
 expectation for the observation of a light supersymmetric 
 Higgs boson at RUNII of
 the Tevatron. Finally, we point out that the BNL constraint,
 specifically the positivity of $\mu$ for a class of models,
   has an important implication for Yukawa unification in grand
   unified models\cite{deboer} and this area is likely to be 
   explored further in the future.

\noindent
\section{Acknowledgments}
This research was supported in part by NSF grant PHY-9901057.

\end{document}